\documentclass[preprint,english,aps,prl,superscriptaddress,12pt]{revtex4-1}
\usepackage[latin9]{inputenc}
\usepackage{amsmath}
\usepackage{amssymb}
\usepackage{graphicx}
\usepackage{esint}
\usepackage{times}
\usepackage{float}
\makeatletter

\usepackage{dcolumn}
\usepackage{bm}
\usepackage{xcolor}

\usepackage{babel}
\makeatother
\usepackage{babel}

\begin{document}

\title{Topological phase transition in layered magnetic compound MnSb$_{2}$Te$_{4}$: Spin-orbit coupling and interlayer coupling dependeces}
\author{Liqin Zhou}
\affiliation{Beijing National Laboratory for Condensed Matter Physics and Institute of physics, Chinese academy of sciences, Beijing 100190, China}
\affiliation{University of Chinese academy of sciences, Beijing 100049, China}

\author{Zhiyun Tan}
\affiliation{School of Physics and Electronic Science, Zunyi Normal University, Zunyi 563006, Guizhou, People's Republic of China}
\affiliation{Beijing National Laboratory for Condensed Matter Physics and Institute of physics, Chinese academy of sciences, Beijing 100190, China}

\author{Dayu Yan}
\affiliation{Beijing National Laboratory for Condensed Matter Physics and Institute of physics, Chinese academy of sciences, Beijing 100190, China}
\affiliation{University of Chinese academy of sciences, Beijing 100049, China}

\author{Zhong Fang}
\affiliation{Beijing National Laboratory for Condensed Matter Physics and Institute of physics, Chinese academy of sciences, Beijing 100190, China}
\affiliation{University of Chinese academy of sciences, Beijing 100049, China}

\author{Youguo Shi}
\email{ygshi@iphy.ac.cn} 
\affiliation{Beijing National Laboratory for Condensed Matter Physics and Institute of physics, Chinese academy of sciences, Beijing 100190, China}
\affiliation{University of Chinese academy of sciences, Beijing 100049, China}

\author{Hongming Weng}
\email{hmweng@iphy.ac.cn} 
\affiliation{Beijing National Laboratory for Condensed Matter Physics and Institute of physics, Chinese academy of sciences, Beijing 100190, China}
\affiliation{University of Chinese academy of sciences, Beijing 100049, China}
\affiliation{Songshan Lake Materials Laboratory, Dongguan, Guangdong 523808, China}

\begin{abstract}
Based on the first-principles calculations and theoretical analysis, we investigate the electronic structures, topological phase transition (TPT) and topological properties of layered magnetic compound MnSb$_{2}$Te$_{4}$. It has the similar crystal and magnetic structure as the magnetic topological insulator MnBi$_{2}$Te$_{4}$. We find that when the spin-orbit coupling (SOC) is considered, the band structure of MnSb$_{2}$Te$_{4}$ in antiferromagnetic (AFM) state has no band inversion at $\Gamma$. This is due to the SOC strength of Sb is less than that of Bi. The band inversion can be realized by increasing the SOC of Sb by 0.3 times, which drives MnSb$_{2}$Te$_{4}$ from a trivial AFM insulator to an AFM topological insulator (TI) or axion insulator. Uniaxial compressive strain along the layer stacking direction is another way to control the band inversion. The interlayer distance shorten by 5$\%$ is needed to drive the similar TPT. For the ferromagnetic (FM) MnSb$_{2}$Te$_{4}$ with experimental crystal structure, it is a normal FM insulator. The band inversion can happen when SOC is enhanced by 0.1 times or the interlayer distance is decreased by more than 1$\%$. Thus, FM MnSb$_{2}$Te$_{4}$ can be tuned to be the simplest type-I Weyl semimetal with only one pair of Weyl nodes on the three-fold rotational axis. These two Weyl nodes are projected onto (1$\bar{1}$0) surface with one Fermi arc connecting them.
\end{abstract}

\maketitle
\section{Introduction}
Topological insulators (TIs) of $Z_2$ classification protected by time-reversal symmetry is characterized by the gapless topological boundary states with Dirac cone like dispersion. It has attracted intensive studies in the field of condensed matter physics~\cite{fu2007topologicala,fu2007topologicalb,hasan2010colloquium,qi2011topological}. When magnetism is induced into TIs, the gapless topological boundary states are expected to be gapped and various exotic topological phenomena will emerge, including topological magnetoelectric effect~\cite{qi2008topological,qi2008fractional}, axion insulator~\cite{li2010dynamical, varnava2018surfaces, yue2019symmetry, xu2019higher} and quantum anomalous hall effect (QAHE)
~\cite{haldane1988model,yu2010quantized,liu2008quantum,nomura2011surface}. The typical three-dimensional (3D) TIs of Bi$_{2}$Se$_{3}$ family have provided a fertile field to host many of these exotic phenomena~\cite{zhang2009topological,chen2009experimental}, especially the QAHE in (Cr, V)-doped (Bi, Sb)$_{2}$Te$_{3}$ thin films ~\cite{yu2010quantized,chang2013experimental,chang2015high}. Recently, magnetic layered material MnBi$_{2}$Te$_{4}$ family has been proposed theoretically to be magnetic TI and many experimental studies have immediately performed to confirm this ~\cite{li2019intrinsic,zhang2019topological,otrokov2019prediction,chowdhury2019prediction,eremeev2017competing,otrokov2019unique,yan2019evolution,swatek2019gapless,murakami2019realization,li2019magnetic,li2019dirac}. It crystalizes in a layered structure with the $R\bar{3}m$ space group (No.166). Each layer is a septuple layer (SL) composed of ``Te-Bi-Te-Mn-Te-Bi-Te" in a triangle lattice \cite{lee2013crystal} and these SLs are stacking through van der Walls interaction. The magnetic ground state of MnBi$_{2}$Te$_{4}$ is layered AFM state. In each SL, the magnetic moments of Mn ions are pointing out of the plan to form ferromagnetic (FM) ordering, and they are antiparallel to those in the neighboring SLs. When the number of SLs various, the thin film of MnBi$_{2}$Te$_{4}$ can change from FM (a single SL) to compensated AFM (even number of SLs) and uncompensated AFM (odd number of SLs). Within external field, it is also possibly be tuned to be FM. Therefore, there have been experimental evidences on Chern insulator, axion insulator, AFM TI and type-II Weyl semimetal (WSM) state realized~\cite{lee2020transport} in MnBi$_{2}$Te$_{4}$ system. The layered crystal structure, layered antiferromagnetic (AFM) configuration and tunable magnetic orders with external magnetic field make it highly attractive in both fundamental research and potential applications.

MnBi$_{2}$Te$_{4}$ can be viewed as intercalating a Mn-Te bilayer into the center of a Bi$_{2}$Te$_{3}$ quintuple layer. There arises a question that whether the topological properties can be preserved when we replace Bi$_{2}$Te$_{3}$ quintuple layer with Sb$_{2}$Te$_{3}$, although Sb$_{2}$Te$_{3}$ is also a 3D strong TI of the same family. However, the spin-orbit coupling (SOC) strength of Sb $5p$ orbitals ($\lambda_{Sb}$=0.4 eV) is far less than that of Bi $6p$ ones ($\lambda_{Bi}$=1.25 eV), we would like to study how the topological states of MnSb$_{2}$Te$_{4}$ are influenced by the SOC and even the interlayer interaction of SLs. In fact, Murakami {\it et al.}~\cite{murakami2019realization} recently have proposed that it is possible to realize FM state in MnSb$_2$Te$_4$ due to the mixing of Mn and Sb sites. They also proposed the FM state might be a type-II WSM based on their calculations. Furthermore,  Shi {\it et al.}~\cite{shi_CPL2020} have observed anomalous Hall effect in MnSb$_2$Te$_4$, supporting the FM or ferrimagnetic order in MnSb$_2$Te$_4$. 


\section{method}
To obtain the electronic structures of MnSb$_{2}$Te$_{4}$, we use the Vienna ab initio simulation package (VASP) with projector augmented wave (PAW) method based on the density functional theory (DFT) \cite{kresse1996efficient,kresse1999ultrasoft}. The Perdew-Burke-Ernzerhof (PBE) exchange-correlation functional with GGA+$U$ method is used to treat the localized $3d$ orbitals of Mn~\cite{perdew1996generalized}. The cutoff energy for the plane-wave basis is set of 520 eV and the $U$ parameter is selected to be 4 eV for Mn $3d$ orbitals. Spin-orbit coupling (SOC) is included self-consistently and we set the local magnetic moment on Mn ions to be along $c$-axis. The Brillouin zone (BZ) integral is implemented on a $\Gamma$ centered grid mesh of 11 $\times$ 11 $\times$ 11 for self-consistent calculations. The surface states and Fermi arc are calculated by using the WannierTools software package based on the maximally localized Wannier functions (MLWF) \cite{wu2018wanniertools,mostofi2008wannier90}. 

Single crystals of MnSb$_2$Te$_4$ were synthesized by using flux method. Starting materials Mn (piece, 99.99$\%$), Sb (grain, 99.9999$\%$) and Te (lump, 99.9999$\%$) were mixed in an Ar-filled glove box at a molar radio of Mn : Sb : Te = 1 : 10 : 16. The mixture was placed in an alumina crucible, which was then sealed in an evacuated quartz tube. The tube was heated to 700 $^{\circ}$C over 10 h and dwelt for 20 h. Then, the tube was slowly cooled down to 630 $^{\circ}$C at a rate of 0.5 $^{\circ}$C/h followed by separating the crystals from the flux by centrifuging. Shiny crystals with large size were obtained on the bottom of the crucible.

\begin{figure}[H]
\centering
\includegraphics[scale=0.35]{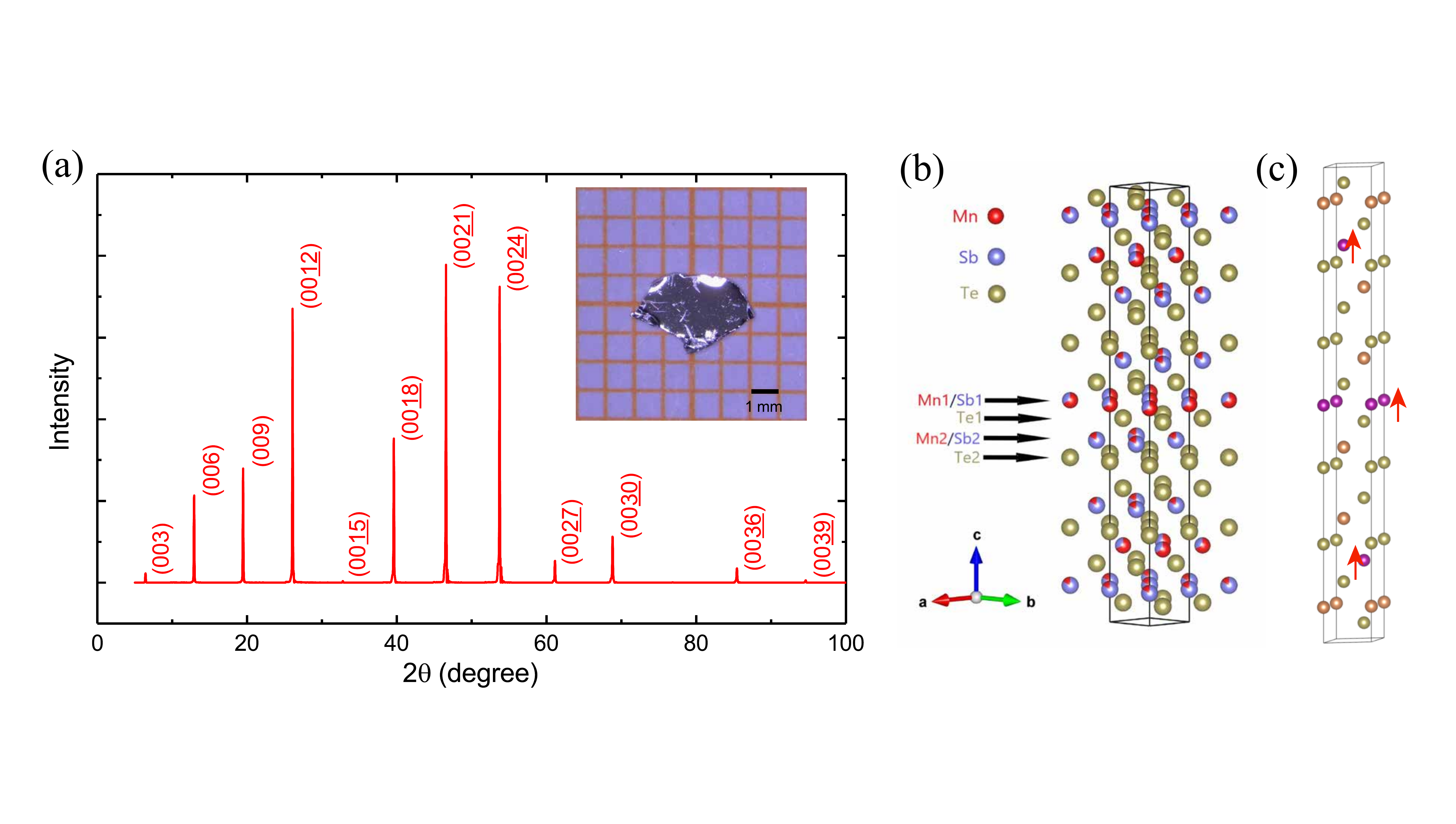}
\caption{(a) The X-ray diffraction pattern of a flat surface of MnSb$_{2}$Te$_{4}$ single crystal. The inset shows a photograph of a typical MnSb$_{2}$Te$_{4}$ single crystal. (b) The schematic crystalline structure of MnSb$_{2}$Te$_{4}$ from experiment with Mn and Sb site mixing. (c) The schematic drawing of the ideal crystal structure for MnSb$_{2}$Te$_{4}$ without mixing of Mn and Sb sites. The arrows around Mn indicate the local magnetic moment on it.
\label{fig0}}
\end{figure}

To investigate the crystalline structure, single-crystal x-ray diffraction (XRD) was carried out on Bruker D8 Venture diffractometer at 293 K using Mo $K\alpha$ radiation ($\lambda$ = 0.71073~\AA). The crystalline structure was refined by full-matrix least-squares method on $F^2$ by using the SHELXL-2016/6 program. The detailed crystallographic parameters are summarized in Table.~\ref{crystr}. The single-crystal XRD study revealed that MnSb$_2$Te$_4$ have the same structure with MnBi$_2$Te$_4$. The lattice parameters of MnSb$_2$Te$_4$ is a = 4.2613~\AA~and c = 41.062~\AA, respectively. Fig.\ref{fig0}(a) shows the XRD patterns of a flat surface of MnSb$_2$Te$_4$ single crystal, where only 00l peaks are detected. A photograph of a typical MnSb$_2$Te$_4$ crystal were shown in the inset of Fig.\ref{fig0}(b), and the back square of 1$\times$1 mm indicates the size of the crystal. Though there is mixing of Mn and Sb sites and it might cause the FM state in MnSb$_2$Te$_4$, we take the ideal crystal structure without such mixing in the calculations. A schematic drawing of the ideal crystal structure based on the experimental one is shown in Fig.\ref{fig0}(c). In space group No. 166, symmetric operations mainly include: three-fold rotation symmetry around the $z$ axis $C_{3z}$, two-fold rotation symmetry around the $x$ and $y$ axis, and inversion symmetry.

\section{Results and discussions}
\subsection{AFM state of MnSb$_{2}$Te$_{4}$} 
\begin{figure}[H]
\centering
\includegraphics[scale=1.0]{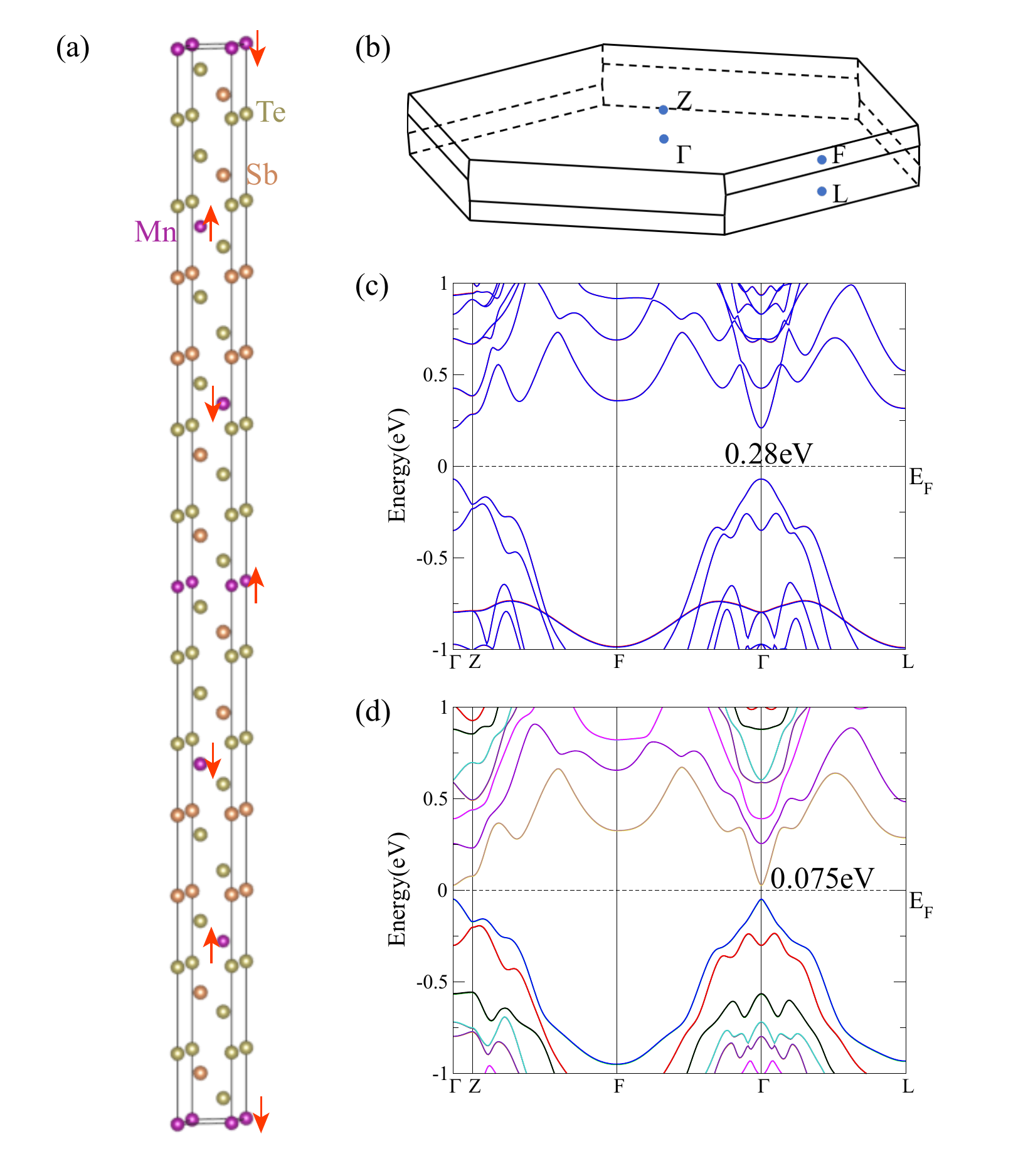}
\caption{Crystal structure and electronic structure of AFM MnSb$_{2}$Te$_{4}$. (a) The unit cell of AFM MnSb$_{2}$Te$_{4}$ and the red arrows represent the spin moment of Mn atom. (b) The first Brillouin zone and four inequivalent TRIM points of MnSb$_{2}$Te$_{4}$. (c) and (d) The band structure of AFM state without (c) and with (d) SOC. \label{fig1}}
\end{figure}

Here, we present and discuss the results of the AFM state firstly. Similar to MnBi$_{2}$Te$_{4}$, the layered antiferromagnetic state has six 
formula units in a conventional unit cell shown in Fig.\ref{fig1}(a). The AFM state has a combined symmetrical operation $S\equiv \Theta \tau_{1/2}$, 
namely time-reversal operator $\Theta$ and the translation operator of half unit cell $\tau_{1/2}$ of the AFM lattice. $\tau_{1/2}$ is formed 
by the nearest neighbor's opposite spin moment of Mn atom layers. According to the GGA+$U$ calculation, we find that the total energy of 
AFM state is lower than that of FM state when SOC is considered, which is the same as that in 
MnBi$_{2}$Te$_{4}$~\cite{li2019intrinsic,zhang2019topological,otrokov2019prediction} and consistent with experimental measurements that MnSb$_{2}$Te$_{4}$ has such layered AFM state. 
Fig.\ref{fig1}(c) and (d) show the band structures of MnSb$_{2}$Te$_{4}$ for AFM state without and with SOC, respectively. 
There is a direct gap of about 0.28 eV at $\Gamma$ point without SOC, but it reduces to about 0.075 eV when SOC is further considered. This 
means SOC can enhance the band inversion as it does in TIs of Bi$_2$Te$_3$ family. Though the magnetism of MnSb$_{2}$Te$_{4}$ 
breaks the time reversal symmetry (TRS), the spatial inversion symmetry $I$ and the combined symmetry $S$ 
are preserved~\cite{mong2010antiferromagnetic,li2019intrinsic,zhang2019topological}. Therefore, the topological invariant $Z_{2}$ protected by $S$ can be obtained through
parity configuration or the evolution of hybrid Wannier function centers (WCCs) of occupied states 
to judge the topological properties of AFM state~\cite{fu2007topologicalb,turner2012quantized,yu2011equivalent}. 
The parities of occupied states at $\Gamma$ point and 
three equivalent F points ($\pi$, $\pi$, 0) indicate that MnSb$_{2}$Te$_{4}$ is not an AFM TI protected by $S$, which is consistent with
the absence of band inversion at $\Gamma$. Since MnBi$_{2}$Te$_{4}$ is an AFM TI and 
the SOC strength of Sb $5p$ orbitals is obviously smaller than that of Bi $6p$, we will manipulate $\lambda_{Sb}$ to  
demonstrate that SOC can drive the topological phase transition.

\begin{figure}[H]
\centering
\includegraphics[scale=0.23]{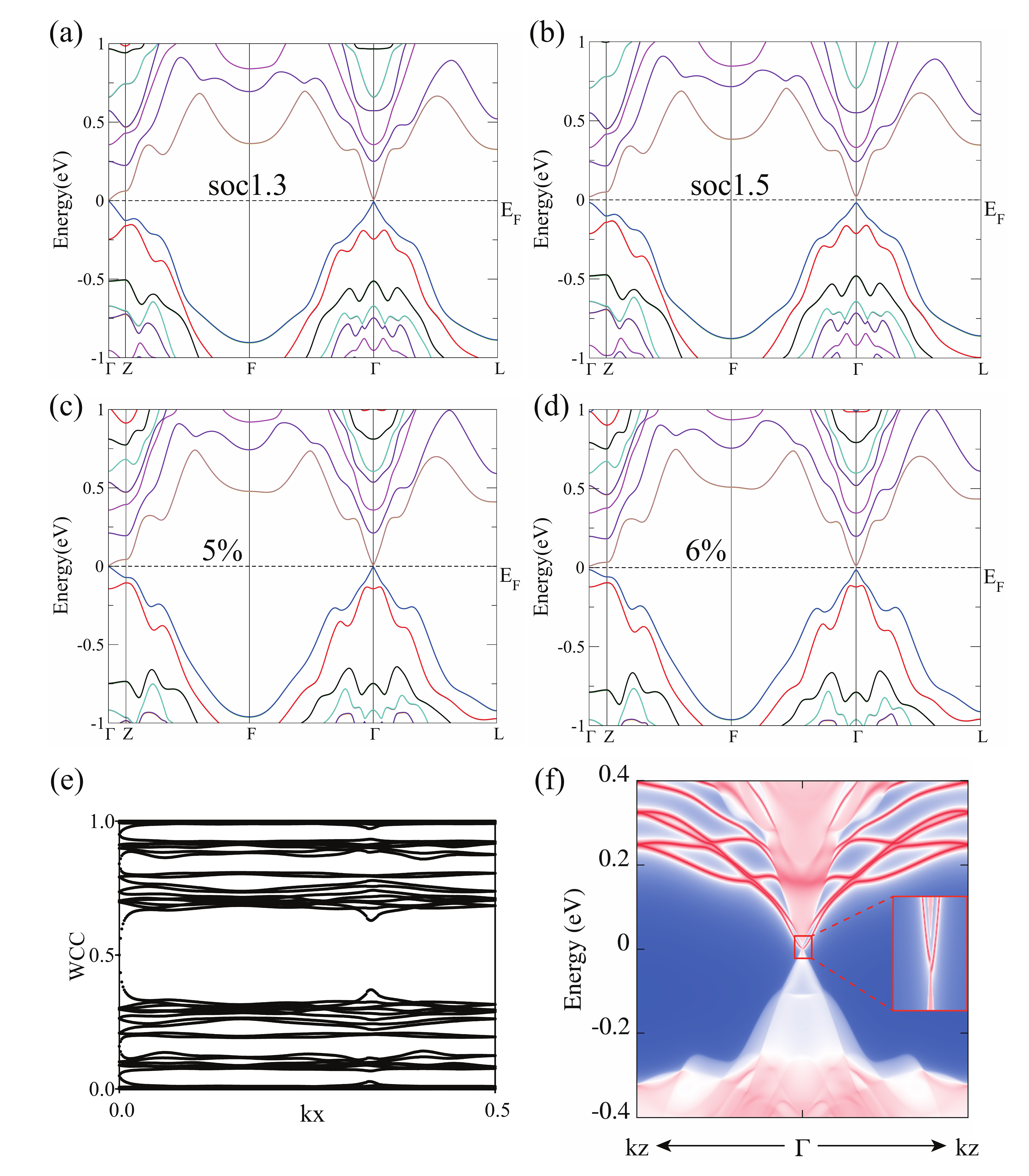}
\caption{The electronic structures of AFM MnSb$_{2}$Te$_{4}$ under different conditions. (a) and (b) Band structures for SOC strength $\lambda_{Sb}$ of Sb changes to 1.3 times and 1.5 times of the original value, respectively. (c) and (d) The band structure with interlayer distance decreased by 5$\%$ and 6$\%$ along $z$ axis with SOC, respectively. (e) Evolution of WCCs in the $k_{z}$ = 0 plane of AFM MnSb$_{2}$Te$_{4}$ under 6$\%$ compressive strain. It implies a nonzero topological invariant. (f) The surface states on the (1$\bar{1}$0) surface. The Dirac cone like surface bands are zoomed out around $\bar{\Gamma}$. \label{fig2}}
\end{figure}

The value of Sb $p$ orbitals, which is parameterized as $\lambda_{Sb}$, is to be tuned in the self-consistent calcualtions.
As shown in Fig.\ref{fig2}(a) and (b), the band structure of MnSb$_{2}$Te$_{4}$ evolves when $\lambda_{Sb}$ increases. Those calculated
when $\lambda_{Sb}$ is 1.3 and 1.5 times of its atomic value $\lambda_{0}$ have been plotted. Obviously, when $\lambda_{Sb}$=1.3$\lambda_{0}$, 
the band gap at $\Gamma$ closes, which is the critical point of topological phase transition. 
When $\lambda_{Sb}$= 1.5$\lambda_{0}$, we find it becomes an AFM TI, the same as MnBi$_{2}$Te$_{4}$, which means that band inversion 
has occurred driven by SOC. According to the critical value of SOC, it is possible to make series of samples MnSb$_x$Bi$_{2-x}$Te$_{4}$ to study 
the topological phase transition.

\par  In addition to directly adjusting the SOC strength $\lambda_{Sb}$ in experimental sample synthesizing through alloying Sb and Bi, we try to simulate the topological phase transition by applying pressure or strain to MnSb$_{2}$Te$_{4}$. This is another usual way to control the physical properties of solids. We simulate the uniaxial compressive strain along $z$-axis by decreasing the interlayer distance among septuple layers along $c$ lattice vector.
Fig.\ref{fig2}(c)-(d) shows the band structures when $c$ is decreased by 5$\%$ to 6$\%$. We find that when $c$ is compressed by about 5$\%$ the band gap closes and it becomes an AFM TI if further compressed. For the case of 6$\%$, the $Z_{2}$ invariant is 1 as shown by the Wilson loop in Fig.~\ref{fig2}(e), indicating that the band inversion has occurred at $\Gamma$. 
We calculate the surface states for the case of 6$\%$ for the (1$\bar{1}$0) surface in Fig.~\ref{fig2}(f), which is a surface that preserves the symmetry $S$=$\Theta \tau_{1/2}$~\cite{li2019intrinsic,zhang2019topological}. It is clear to see that there is a Dirac cone like topological surface states in the band gap connecting the conduction and valence band, respectively, though the band gap is quite small that the lower Dirac cone is buried beneath the bulk states. 
On the contrary, we can only see a trivial gapped surface state on the (111) surface, which does not preserve $S$ symmetry.

\subsection{FM state of MnSb$_{2}$Te$_{4}$} 
\begin{figure}[H]
\centering
\includegraphics[scale=0.3]{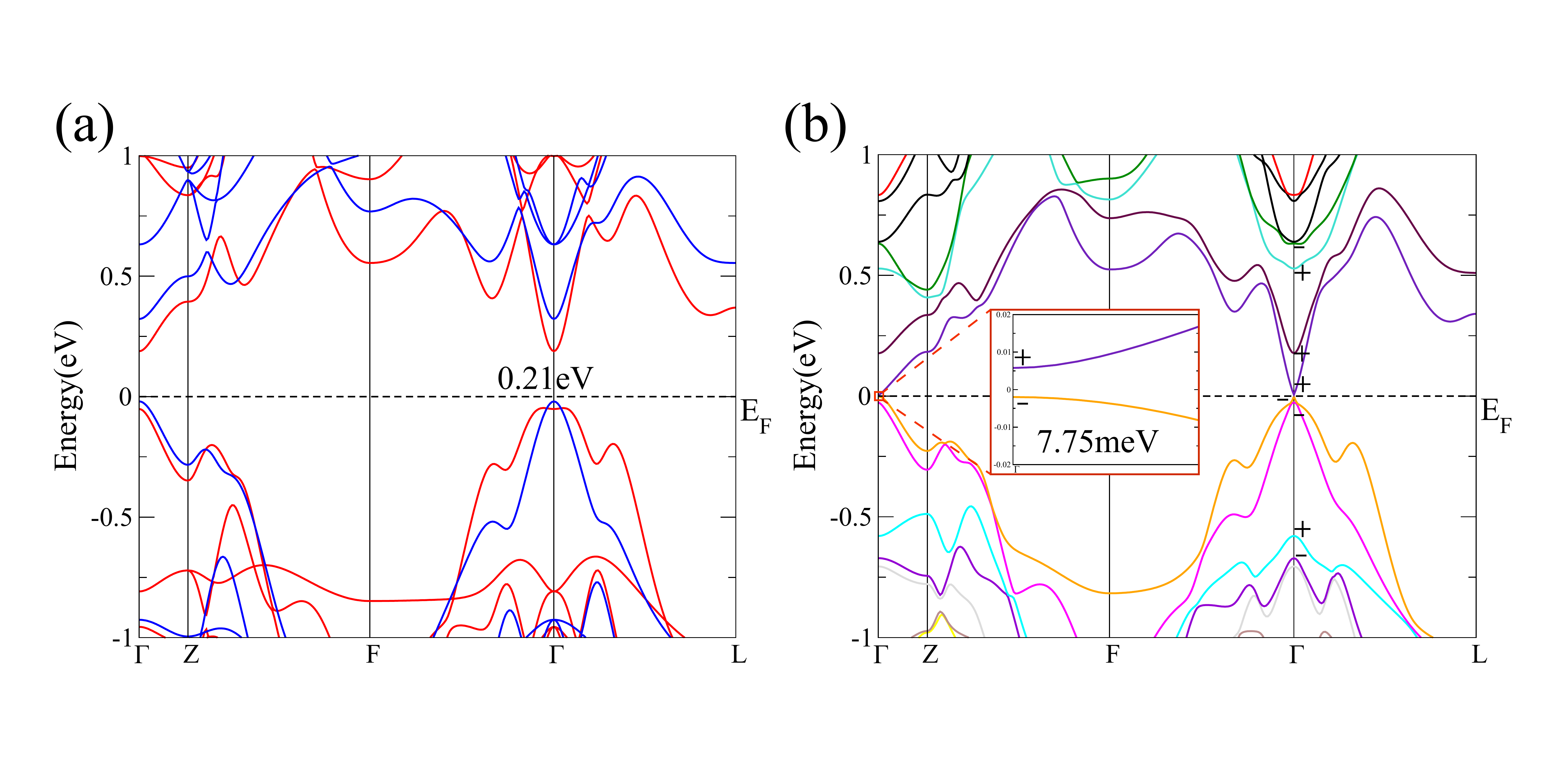}
\caption{The electronic structures of FM MnSb$_{2}$Te$_{4}$. (a) and (b) The band structures of FM state without (a) and with (b) SOC (red bands indicate spin up and blue bands indicate spin down in a). \label{fig4}}
\end{figure}

Now we discuss the results of the FM MnSb$_{2}$Te$_{4}$. The electronic structures of FM MnSb$_{2}$Te$_{4}$ calculated without and with SOC are shown in Fig.\ref{fig4}(a) and (b), respectively. There is a direct band gap about 0.21 eV between spin-up and spin-down bands at $\Gamma$ point when SOC is not included. When SOC is taken into account, the band gap is reduced very much to be about 7.75 meV at the $\Gamma$ point. This is consistent with the observation in the above that SOC will enhance the band inversion. 

\begin{figure}[H]
\centering
\includegraphics[scale=0.23]{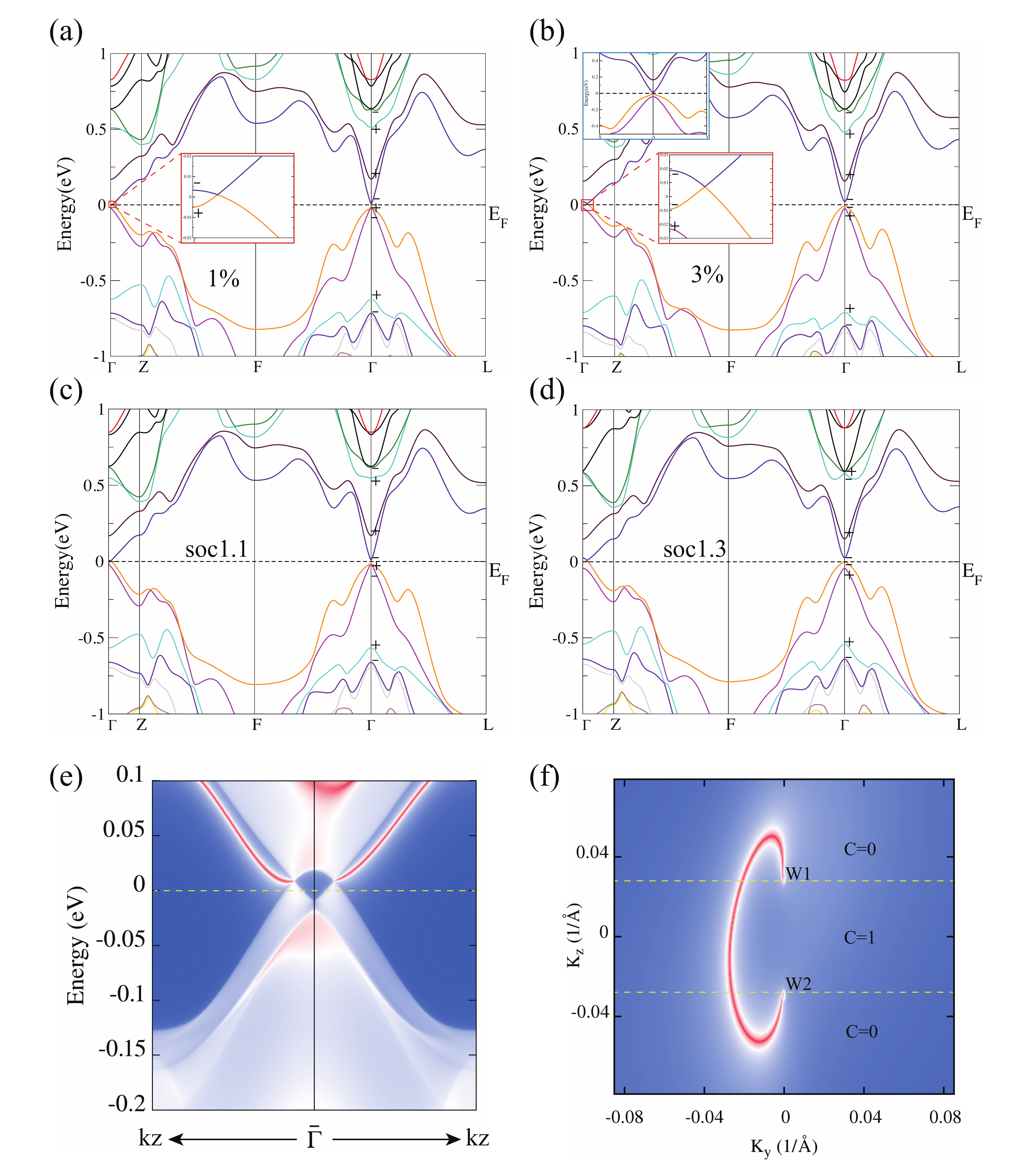}
\caption{The electronic structures of FM MnSb$_{2}$Te$_{4}$ under different conditions. (a) and (b) The band structures with interlayer distances decreased by 1$\%$ and 3$\%$ with SOC, respectively. (The insert of blue border shows the bands along the $k_{x}$ and $k_{y}$ axis through the Weyl point in b, which indicates a type-I Weyl point) (c) and (d) Band structures for SOC strength $\lambda_{Sb}$ of Sb changes to 1.1 and 1.3 times of the original value, respectively. (e) Surface state for case with 3$\%$ compressive strain on the (1$\bar{1}$0) surfaces. There are two Weyl nodes along the $k_{z}$ direction. (f) Fermi arc connecting the projections of the Weyl nodes W1 and W2 at 0.007eV. \label{fig5}}
\end{figure}

To check the topological quantum of FM MnSb$_{2}$Te$_{4}$, we calculate the topological indice $z_4$ using the parity eigenvalue $p_n(\Lambda)$ of occupied states $n$ at eight time-reversal invariant momenta (TRIM) $\Lambda$ since the inversion symmetry is kept~\cite{turner2012quantized, xu2020high}. $z_4$ is defined as 
\begin{equation}
\label{eq:e1}
\begin{aligned}
z_{4}=\sum_{\Lambda \in TRIM} \sum_{n \in occ}\frac{1+p_n(\Lambda)}{2} ~ mod ~ 4.
\end{aligned}
\end{equation}
If $z_4$ = 1, 3, it means a Weyl semimetal (WSM) phase, where an odd number of Weyl nodes exist in half of the BZ. If $z_4$ = 2, it indicates an axion insulator. The above AFM TI state after tuning can also be looked as an axion insulator.
We find that the FM MnSb$_{2}$Te$_{4}$ with experimental lattice structure
is a trivial insulator with full gap throughout the whole BZ with $z_4$=0. The parity eigenvalues of the bands around the Fermi level at $\Gamma$ have been indicated with ``+" or ``-" in Fig.~\ref{fig4}(b) since only the bands at $\Gamma$ will have band inversion during the tuning. 
In order to realize WSM state, we pressurized the FM structure of MnSb$_{2}$Te$_{4}$ along $c$ axis as done in AFM case. Fig.\ref{fig5}(a) and (b) show the calculated bands of FM structure compressed by 1$\%$ and 3$\%$, respectively. Through the parity configuration of the bands at $\Gamma$, one can immediately find that $z_4$ changes from 0 to 1, indicating there is odd number of Weyl nodes in half of the BZ. Comparing with the AFM state, 1$\%$ compression is enough to drive the topological phase transition and generates cross points, i.e. Weyl nodes. We have searched that the Weyl nodes are on the $\Gamma$-Z path, i.e. on the $C_{3z}$ rotation axis. The $C_{3z}$ symmetry is preserved in either AFM or FM states we studied with Mn local magnetic moment pointing parallel to the axis. The inversion symmetry relates the Weyl node in $k_{z}$$>$0 BZ with its pair partner of opposite chirality in $k_{z}$$<$0 BZ. If the Weyl nodes are away from the $C_{3z}$ rotation axis, there will be at least three (odd number) pairs of them in the whole BZ. 
These Weyl nodes are type-I with upright cones shown in the insets of Fig.~\ref{fig5}(a) and (b). We choose the structure under 3$\%$ compressive strain to show its Weyl nodes, surface states and Fermi arc. As shown in Fig.\ref{fig5}(e), we calculate the (1$\bar{1}$0) surface. $k_{z}$ is along the projection line of path $\Gamma$-Z of 3D BZ. The two solid Dirac cones close to $\bar{\Gamma}$ near Fermi level are the projections of two Weyl nodes. Their chirality is 1 and -1, respectively, and the energy is very close to the Fermi level ($E_{arc}$ $\approx$ 0.007eV). 
There are two a very clear surface bands connecting the projections of the two Weyl nodes as depicted in Fig.\ref{fig5}(e), and they lead to the Fermi arc connecting the Weyl nodes is also clearly visible in Fig.\ref{fig5}(f). These are obvious and typical characterizations of type-I WSM \cite{weng2015weyl,xu2015discovery,lv2015experimental}. It is known that the insulating electronic state constrained within the two-dimensional plane perpendicular to $k_z$ will have different Chern number $C$ when the plane locates between the two Weyl nodes or out of them as indicated in Fig.~\ref{fig5}(f)~\cite{xu2011chern, weng2015quantum}. This distribution of $C$ is consistent with the number of crossing points between Fermi arc and a reference line in horizontal direction, namely even times (zero or two) in region with $C$=0 and odd times (one) in region with $C$=1.
\par We also adjust the SOC strength of Sb for the FM state with 1.1 and 1.3 times of the original SOC strength $\lambda_{0}$ and the bands are shown in Fig.~\ref{fig5}(c) and (d). Compared with AFM state, it is more prone to emerge topological phase transition for FM state, and 1.1 times of can result in the occurrence of band inversion and Weyl nodes. 

\section{Summary}  
We mainly investigate the electronic structures, band topology and the surface states for the AFM and FM state of layered magnetic material MnSb$_{2}$Te$_{4}$, which has been synthesized experimentally. The dependences of these on the SOC strength and the uniaxial strain have been simulated by first-principles calculations. We find that in the AFM state, the SOC strength of Sb is too small and MnSb$_{2}$Te$_{4}$ is not an AFM TI. However, the band inversion can be realized when the interlayer distance is decreased by more than 5$\%$, or the SOC strength $\lambda_{Sb}$ of Sb is increased to more than its 1.3 times. Both can drive topological phase transition and lead MnSb$_{2}$Te$_{4}$ to AFM TI. In the FM state, MnSb$_{2}$Te$_{4}$ is not a WSM in its experimental structure. The compressive strain decreasing the interlayer distance by about 1$\%$ or increasing $\lambda_{Sb}$ by about 1.1 times can realize the band inversion, and drive the topological phase transition to WSM. The Weyl nodes appear on the $\Gamma$-Z path and they are type-I upright Weyl cone. 

\begin{acknowledgments}
This work was supported by the National Natural Science Foundation of China  (11504117, 11674369, 11925408, 11974395, 11921004).
Z.T. acknowledges the Foundation of Guizhou Science and Technology Department under Grant No. QKH-LHZ[2017]7091.
H.W. acknowledges support from the National Key Research and Development Program of China (Grant Nos. 2016YFA0300600 and 2018YFA0305700), the K. C. Wong Education Foundation (GJTD-2018-01), the Beijing Natural Science Foundation (Z180008), and the Beijing Municipal Science and Technology Commission (Z181100004218001).
\end{acknowledgments}

\bibliographystyle{unsrt}
\bibliography{ref}

\begin{table}[H]
\centering
\caption{Crystallographic data of MnSb$_{2}$Te$_{4}$.}
\setlength{\tabcolsep}{4mm}{
\begin{tabular}{ccccccc}
atom&x&y&z&Occ.&Wyckoff Position&Sym.\\
\hline
Mn1&0.00000&1.00000&0.50000&0.680&3b&-3m\\
Sb1&0.00000&1.00000&0.50000&0.320&3b&-3m\\
Te1&0.33333&0.66667&0.45887&1&6c&3m\\
Mn2&0.66667&0.33333&0.40824&0.172&6c&3m\\
Sb2&0.66667&0.33333&0.40824&0.828&6c&3m\\
Te2&1.00000&0.00000&0.36864&1&6c&3m\\
\hline
\end{tabular}}
\label{crystr}
\end{table}

\newpage

\end{document}